\documentclass[prb,twocolumn,amssymb,amsmath,showpacs,eqsecnum]{revtex4}

\usepackage{graphicx}
\usepackage{dcolumn}
\usepackage{bm}

\begin{document}

\title{Manifestation of the Roughness-Square-Gradient Scattering \\
in Surface-Corrugated Waveguides}

\author{F.~M.~Izrailev}
\email{izrailev@venus.ifuap.buap.mx}

\affiliation{Instituto de F\'{\i}sica, Universidad Aut\'{o}noma de Puebla, \\
         Apartado Postal J-48, Puebla, Pue., 72570, M\'{e}xico}

\author{N.~M.~Makarov}
\email{makarov@siu.buap.mx}

\affiliation{Instituto de Ciencias, Universidad Aut\'{o}noma
         de Puebla, \\ Priv. 17 Norte No. 3417, Col. San Miguel
         Hueyotlipan, Puebla, Pue., 72050, M\'{e}xico}

\author{M.~Rend\'{o}n}
\email{mrendon@venus.ifuap.buap.mx}

\affiliation{Facultad de Ciencias de la Electr\'onica,
         Universidad Aut\'{o}noma de Puebla,
         Puebla, Pue., 72570, M\'{e}xico}

\date{\today}

\begin{abstract}
We study a new mechanism of wave/electron scattering in multi-mode
surface-corrugated waveguides/wires. This mechanism is due to
specific square-gradient terms in an effective Hamiltonian
describing the surface scattering, that were neglected in all
previous studies. With a careful analysis of the role of roughness
slopes in a surface profile, we show that these terms strongly
contribute to the expression for the inverse attenuation length
(mean free path), provided the correlation length of corrugations
is relatively small. The analytical results are illustrated by
numerical data.
\end{abstract}

\pacs{42.25.Dd; 73.21.Hb; 73.23.-b;73.50.Bk;73.63.Nm}

\maketitle

\section{Introduction}
\label{sec-Intro}

The subject of wave transport through guiding surface-disordered
systems is of great importance in various physical applications.
Recently, this topic has attracted even more attention due to a
burst of developments in nano-science where frequently one deals
with devices or structures (quantum wires, leads, etc.) whose
surface's irregularities become more important than those in the
bulk. Other situations for either electromagnetic or acoustic
waves (remote sensing, photonic and acoustic devices, optical thin
films, etc.) are analogous. Therefore, in what follows we do not
make any distinction between electromagnetic/electron or any other
type of waves.

In spite of an extensive research of unitary wave scattering from
rough surfaces, the problem of transport in the waveguides with
such profiles still remains open. This problem is a great
theoretical challenge since it deals with the multiple scattering
of a wave from lateral walls. As a result of this scattering, the
unperturbed longitudinal wave number $k_n$ of an $n$-th
propagating normal mode changes, $k_n\to k_n+\delta k_n$, due to a
complex amount $\delta k_n$,
\begin{equation}\label{k+dk}
\delta k_n=\gamma_n+i(2L_n)^{-1}.
\end{equation}
The real part $\gamma_n$ is responsible for the roughness-induced
correction to the phase velocity, while $L_n$ is the electron {\it
total mean free path}, {\it scattering length} or {\it attenuation
length} of a given mode. As is known, the shift $\gamma_n$ does
not change the static transport of a disordered systems.
Therefore, our further analysis shall be focused only on the
attenuation length $L_n$.

Evidently, statistical properties of a rough surface profile
$\xi(x)$ give a strong impact on the scattering process. However,
a proper incorporation of these properties into the theory
accounting for the guided and scattered waves, is not a trivial
task. Many approaches have been proposed in connection with the
surface scattering (see, e.g., Refs.~\onlinecite{BFb79,
RytKrTat89,VoronB94,SMFY98,SFMY99,Konrady74,McGM84,TJM86,TrAsh88,
IsPuzFuks9091,Tatar93,KMYa91,MakMorYam95,MT9801,MeyStep949597989900,
LunKyReiKr96,BratRash96,LunMenIz01} and references therein). One
of the main tools to treat this problem is a reduction of the {\it
surface} scattering to the {\it bulk} one in such a way that the
latter can be formally described by an effective Hamiltonian
$\widehat H$,
\begin{equation}\label{H0-U}
\widehat H^{(0)}\to \widehat H=\widehat H^{(0)}+\widehat U,
\end{equation}
with flat boundaries, however, with a complicated potential
$\widehat U$. To the best of our knowledge, originally the idea of
this approach was discussed by Migdal~\cite{Mig77}. After, it was
frequently used in the theories of classical and quantum wave
scattering, see, e.g.,
Refs.~\onlinecite{Konrady74,TJM86,TrAsh88,BratRash96,
MeyStep949597989900,LunKyReiKr96,IsPuzFuks9091,MT9801,LunMenIz01}.
One should stress that in the majority of
them~\cite{Konrady74,TJM86,TrAsh88,
MeyStep949597989900,LunKyReiKr96,BratRash96} the study was
restricted to the lowest order in the root-mean-square roughness
height $\sigma$. Other methods \cite{IsPuzFuks9091,MT9801,
LunMenIz01,VoronB94,KMYa91,MakMorYam95,Tatar93} were mainly based
on a principal assumption that the surface roughness is
sufficiently smooth.

In this paper we present analytical results demonstrating a new
scattering mechanism missed in previous studies of the surface
scattering. This mechanism is due to specific {\it
square-gradient} (SG) terms that are proportional to
$\sigma^2{\xi'}^2(x)$ in the potential $\widehat U$. We argue that
the discovered scattering mechanism is substantially different
from the known ones associated with the {\it roughness amplitude}
and {\it roughness gradient} of the surface.

The two last mechanisms have been already studied (see, e.g.
Refs.~\onlinecite{BFb79,MeyStep949597989900,LunKyReiKr96}). Their
main contributions were shown to depend on the terms containing
the quantities $\sigma\xi(x)$ and $\sigma\xi'(x)$, respectively.
In the performed analysis the square-gradient terms related with
the new scattering mechanism were discarded due their seemingly
small contribution. Indeed, the square-gradient terms are formally
proportional to $\sigma^2$, similar to other terms that arise in
the next (second-order) approximation in the amplitude and
gradient roughness. However, we have found that the
square-gradient terms have a very strong dependence on the
roughness correlation length $R_c$, in contrast with those terms
appearing in the second-order approximation. Specifically, with a
decrease of $R_c$ the square-gradient terms in the expression for
the mean free path compete with the standard terms proportional to
$\sigma\xi(x)$ and $\sigma\xi'(x)$.

One should stress that our approach is restricted by the first
order approximation, with a careful analysis of all terms that may
be important in this approximation, and taking into account an
unexpectedly strong influence of the square-gradient terms.  Our
goal is to study the new scattering mechanism and establish the
conditions when it should not be neglected. For this, we derive a
correct expression for the attenuation length $L_n$ that
incorporates the contribution of the above mechanism. One should
emphasize that the approach we use does not assume any special
restrictions to the model parameters (such as the smoothness of
surface profiles) except for general conditions of {\it weak
scattering}.

The paper is organized as follows. In Sec.~\ref{sec-problem} we
formulate the problem and discuss the coordinate transformation
used to represent the surface scattering as the bulk one. Our
approach involves an average Green's function whose longitudinal
wave number is a modification of the unperturbed one. Also, all
expressions corresponding to the unperturbed problem are given and
discussed in this Section. In Sec.~\ref{sec-1RB-Dyson} we derive a
Dyson-type integral equation for the exact Green's function. From
the exact expression for the scattering potential we design two
Hermitian random operators. The first one is associated with both
the {\it amplitude} (AS) and the {\it gradient scattering} (GS)
mechanism, while the second one is associated only with the {\it
square-gradient scattering} (SGS) mechanism. The former operator
gives rise to the {\it roughness-height power} (RHP) spectrum
$W(k_x)$, while the latter to the {\it roughness-square-gradient
power} (RSGP) spectrum $T(k_x)$.

In Sec.~\ref{sec-1RB-AverageGF} we obtain the averaged Green's
function by applying a perturbative method with respect to the
above operators. Since in this work we restrict ourselves to the
analysis of the attenuation length $L_n$ of the $n$-th conducting
mode, we focus our attention to the imaginary part of the proper
self-energy. In Sec.~\ref{sec-1RB-Ln-An}, in correspondence with a
clear independence between AS+GS and SGS mechanisms, we develop a
natural approach to define two attenuation lengths, the well known
one, $L_n^{(1)}$, and the new SGS length $L_n^{(2)}$. These two
lengths are related to the RHP and the RSGP spectrum, and, as a
consequence, they are characterized by $\sigma^2$ and $\sigma^4$
dependencies, respectively. We study their interplay for two limit
cases of the roughness surface, for small-scale and large-scale
roughness. We discuss here the main result according to which the
larger roughness slope $\sigma/R_c$, the larger contribution of
the SGS mechanism. We also present a numerical analysis assuming
that the surface profile $\xi(x)$ has the standard Gaussian binary
correlator. Finally, in Section \ref{Sec-Conclusion} we outline
our conclusions. A short presentation of main results can be found
in Ref.~\onlinecite{IzMakRen05}.

\section{Problem formulation}
\label{sec-problem}

In what follows, we consider an open plane waveguide (or
conducting quasi-one-dimensional wire) of average width $d$,
stretched along the $x$-axis. For simplicity, one (lower) surface
of the waveguide is assumed to be flat, $z=0$, while the other
(upper) surface has a rough profile, $z=d+\sigma\xi(x)$, with
$\sigma$ as the root-mean-square roughness height. In other words,
the waveguide occupies the region
\begin{equation}\label{1RB}
-\infty\leq x\leq\infty,\qquad\qquad 0\leq z\leq w(x)
\end{equation}
of the $(x,z)$-plane. Here the fluctuating wire width $w(x)$ is
defined by
\begin{equation}\label{1RB-d(x)}
w(x)=d+\sigma\xi(x),\qquad\qquad \langle w(x)\rangle=d.
\end{equation}

The random function $\xi(x)$ describes the roughness of the upper
boundary. It is assumed to be a statistically homogeneous and
isotropic {\it Gaussian} random process with the zero mean and
unit variance,
\begin{equation}\label{ksi}
\langle\xi(x)\rangle=0,\quad \langle\xi^2(x)\rangle=1,\quad
\langle\xi(x)\xi(x')\rangle={\cal W}(x-x').
\end{equation}
Here the angular brackets stand for statistical averaging over
different realizations of the surface profile $\xi(x)$. We also
assume that its binary correlator ${\cal W}(x)$ decreases on the
scale $R_c$ and is normalized to one, ${\cal W}(0)=1$.

The roughness-height power (RHP) spectrum $W(k_x)$ is defined by
\begin{equation}\label{FT-W}
W(k_x)=\int_{-\infty}^{\infty}dx\exp(-ik_xx)\,{\cal W}(x).
\end{equation}
Since ${\cal W}(x)$ is an even function of $x$, its Fourier
transform (\ref{FT-W}) is even, real and non-negative function of
$k_x$. The RHP spectrum has maximum at $k_x=0$ with $W(0)\sim
R_c$, and decreases on the scale $R_c^{-1}$.

In order to analyze the surface scattering problem we shall employ
the method of retarded Green's function ${\cal G}(x,x';z,z')$.
Specifically, we start with the Dirichlet boundary-value problem
\begin{subequations}\label{1RB-GFP1}
\begin{eqnarray}
&&\left(\frac{\partial^2}{\partial x^2}+
\frac{\partial^2}{\partial z^2}+k^2\right){\cal G}(x,x';z,z')
\nonumber\\[6pt]
&&=\delta(x-x')\delta(z-z'),\label{1RB-Geq1}\\[6pt]
&&{\cal G}(x,x';z=0,z')=0,\nonumber\\
&&{\cal G}(x,x';z=d+\sigma \xi(x),z')=0.\label{1RB-BC1}
\end{eqnarray}
\end{subequations}
Here $\delta(x)$ and $\delta(z)$ are the Dirac delta-functions.
The wave number $k$ is equal to $\omega/c$ for an
electromagnetic wave of the frequency $\omega$ and TE polarization,
propagating through a waveguide with perfectly conducting walls.
For an electron quantum wire, $k$ is the Fermi wave number within
the isotropic Fermi-liquid model.

\subsection{Coordinate transformation}
\label{subsec-Gen}

The equation (\ref{1RB-Geq1}) does not contain any scattering
potential. In contrast with the bulk scattering, the
electromagnetic/electron waves experience a perturbation due to
scattering from the upper wall, therefore, the perturbation is
hidden in the boundary condition (\ref{1RB-BC1}). In order to
formally describe the surface scattering as a bulk one, we perform
the canonical transformation to new coordinates,
\begin{subequations}\label{1RB-CoorTr}
\begin{eqnarray}
x_{new}&=& x_{old},\\
z_{new}&=& \frac{d}{w(x)}z_{old}= \frac{d}{d+\sigma \xi(x)} z_{old},
\end{eqnarray}
\end{subequations}
in which both waveguide surfaces are flat. Correspondingly, we
introduce the canonically conjugate Green's function (below for
convenience we drop the subscript ``{\it new}'' for $x$ and $z$),
\begin{equation}
{\cal G}_{new}(x,x';z,z')=\frac{\sqrt{w(x)w(x')}}{d}\,
{\cal G}_{old}(x,x';z,z').
\end{equation}
As a result, we arrive at the equivalent boundary-value problem
governed by the equation
\begin{subequations}\label{1RB-GFP4}
\begin{eqnarray}
&&\Bigg(\frac{\partial^2}{\partial x^2}+
\frac{\partial^2}{\partial z^2} +k^2\Bigg)
\,{\cal G}(x,x';z,z')\nonumber\\[6pt]
&&-\widehat{U}(x,z)\,{\cal G}(x,x';z,z')
=\delta(x-x')\delta(z-z'),\qquad\label{1RB-Geq4}\\[6pt]
&&{\cal G}(x,x';z=0,z')=0,\nonumber\\
&&{\cal G}(x,x';z=d,z')=0.
\label{1RB-BC4}
\end{eqnarray}
\end{subequations}
Here the effective {\it surface scattering potential}
$\widehat{U}(x,z)$ is given by the following expression,
\begin{eqnarray} \label{1RB-U}
\widehat{U}(x,z)&=&\bigg[1-\frac{d^2}{w^2(x)}\bigg]
\frac{\partial^2}{\partial z^2}\nonumber\\[6pt]
&+&\frac{\sigma}{w(x)}\left[\xi''(x)+ 2\xi'(x)
\frac{\partial}{\partial x}\right]
\left[\frac{1}{2}+z\frac{\partial}{\partial z}\right]\nonumber\\[6pt]
&-&\frac{\sigma^2{\xi'}^2(x)}{w^2(x)}\left[\frac{3}{4}+3z
\frac{\partial}{\partial z}+z^2\frac{\partial^2}{{\partial
z}^2}\right].
\end{eqnarray}
Note that the prime over the function $\xi(x)$ denotes a
derivative with respect to $x$. It is important to stress that
Eqs.~(\ref{1RB-GFP4})--(\ref{1RB-U}) are {\it exact} and valid for
any form of the surface profile $\xi(x)$.

\subsection{Unperturbed Green's function}
\label{subsec-UnpProb}

The unperturbed Green's function ${\cal G}_0(|x-x'|;z,z')$ obeys
the boundary-value problem (\ref{1RB-GFP4}) with
$\widehat{U}(x,z)=0$ (when $\sigma=0$, and therefore, $w(x)=d$).
It is determined as follows,
\begin{subequations}\label{G0}
\begin{eqnarray}
&&{\cal G}_0(|x-x'|;z,z')=\int_{-\infty}^{\infty}\frac{dk_xdk_x'}{(2\pi)^2}\,
\exp\left[i(k_xx-k_x'x')\right]\nonumber\\[6pt]
&&\times g_0(k_x,k_x')\,\frac{\sin(k_zz)\sin(k_z'z')}{k_zk_z'}
\label{G0-pole-kk'}\\[6pt]
&&=\sum_{n=1}^{N_d}\sin\left(\frac{\pi nz}{d}\right)
\sin\left(\frac{\pi nz'}{d}\right)
\frac{\exp\left(ik_n|x-x'|\right)}{ik_nd}.\qquad\quad
\label{G0-mode}
\end{eqnarray}
\end{subequations}
Here $k_x$ and $k_z$ are the lengthwise and transverse wave
numbers,
\begin{equation}\label{kz}
k_z=k_z(k_x)=\sqrt{k^2-k_x^2},\qquad k_z'=k_z(k_x').
\end{equation}
Their unperturbed eigenvalues are $k_n$,
\begin{equation}\label{kn}
k_n=\sqrt{k^2-(\pi n/d)^2},\qquad n=1,2,3,\ldots,N_d,
\end{equation}
and $\pi n/d$ respectively. The total number $N_d$ of propagating
waveguide modes (conducting channels that have real values of
$k_n$) is determined by the integer part $[\ldots]$ of the ratio
$kd/\pi$,
\begin{equation}\label{Nd}
N_d=[kd/\pi].
\end{equation}

The first expression (\ref{G0-pole-kk'}) gives the unperturbed
Green's function in the {\it pole representation}
where $g_0(k_x,k_x')$ is called the {\it pole factor},
\begin{subequations}\label{g0}
\begin{eqnarray}
&&g_0(k_x,k_x')=2\pi\delta(k_x-k_x')\,g_0(k_x),
\label{g0-kk'}\\[6pt]
&&g_0(k_x)=k_z\cot(k_zd)\,\to
\label{g0-cot}\\[6pt]
&&\sum_{n=1}^{N_d}\frac{(\pi n/d)^2}{k_nd}
\left(\frac{1}{k_x+k_n+i0}-\frac{1}{k_x-k_n-i0}\right).
\qquad\quad\label{g0-pf}
\end{eqnarray}
\end{subequations}
The evaluation of the integrals in Eq.~(\ref{G0-pole-kk'}) over
the poles of $g_0(k_x,k_x')$, results in the mode representation
(\ref{G0-mode}) for the unperturbed Green's function.

\section{Dyson Equation}
\label{sec-1RB-Dyson}

With the use of the Green's theorem, it can be easily shown that
the boundary-value problem (\ref{1RB-GFP4}) is directly reduced to
the following {\it exact} Dyson-type integral equation
\begin{eqnarray}
&&{\cal G}(x,x';z,z')={\cal G}_0(|x-x'|;z,z')+
\int_{-\infty}^{\infty}dx_1\int_0^ddz_1 \nonumber\\[6pt]
&&{\cal G}_0(|x-x_1|;z,z_1)\,\widehat{U}(x_1,z_1)\,{\cal
G}(x_1,x';z_1,z').\qquad\label{1RB-xDE}
\end{eqnarray}
This equation relates the perturbed by surface disorder Green's
function ${\cal G}(x,x';z,z')$ to the Green's function ${\cal
G}_0(|x-x'|;z,z')$ of the waveguide with perfectly flat
boundaries.

Similarly to the pole representation (\ref{G0-pole-kk'}) for the
unperturbed Green's function, let us seek ${\cal G}(x,x';z,z')$ in
the form
\begin{eqnarray}
&&{\cal G}(x,x';z,z')=
\int_{-\infty}^{\infty}\frac{dk_xdk_x'}{(2\pi)^2}
\exp\left[i(k_xx-k_x'x')\right]
\nonumber\\[6pt]
&&\times g(k_x,k_x')\,\frac{\sin(k_zz)\sin(k_z'z')}{k_zk_z'}.
\label{G-pole}
\end{eqnarray}

With this method the problem of deriving the Green's function
${\cal G}(x,x';z,z')$ is reduced to obtain its pole factor
$g(k_x,k_x')$. To this end, we substitute the pole representations
(\ref{G0-pole-kk'}) and (\ref{G-pole}) into Eq.~(\ref{1RB-xDE})
and get the non-local Dyson equation in the $k_x$-representation,
\begin{eqnarray}
&&g(k_x,k_x')=g_0(k_x,k_x')\nonumber\\[6pt]
&&+\int_{-\infty}^{\infty}\frac{dq_xdq_x'}{(2\pi)^2}g_0(k_x,q_x)
\Xi(q_x,q_x')g(q_x',k_x').\qquad\qquad\label{g-DE}
\end{eqnarray}
In this representation the effective surface scattering potential
$\Xi(k_x,k'_x)$ is
\begin{eqnarray}
&&\Xi(k_x,k'_x)=\int_{-\infty}^{\infty}dx_1\int_0^ddz_1\,
\exp\left(-ik_xx_1\right)\frac{\sin(k_zz_1)}{k_z}
\nonumber\\[6pt]
&&\times \widehat{U}(x_1,z_1)\;
\exp\left(ik'_xx_1\right)\frac{\sin(k'_zz_1)}{k'_z}\,.
\label{1RB-Xi-exact}
\end{eqnarray}

Thus, we obtained the Dyson-type integral equation (\ref{g-DE})
with an {\it exact} expression (\ref{1RB-Xi-exact}) for the kernel
$\Xi(k_x,k'_x)$. After the substitution of the expression
(\ref{1RB-U}) for $\widehat{U}$, we realize that the kernel
consists of three groups of terms. The first one has the factor
$(1-d^2/w^2(x_1))$ while the second and third group has,
respectively, the factor $\sigma/w(x_1)$ and $\sigma^2/w^2(x_1)$.
One needs to note that while the kernel $\Xi(k_x,k'_x)$ is
Hermitian in the whole, its latter two parts are non-Hermitian
individually. To have each part of $\Xi(k_x,k'_x)$ Hermitian, we
perform the integration by parts for the term containing
$\xi''(x_1)$ in the second group. After that we come to the final
{\it exact} form for the perturbation potential,
\begin{eqnarray}
&&\Xi(k_x,k'_x)=\int_{-\infty}^{\infty}dx_1\int_0^ddz_1\,
\exp\left(-ik_xx_1\right) \frac{\sin(k_zz_1)}{k_z}\nonumber\\[6pt]
&&\times \left\{\left[1-\frac{d^2}{w^2(x_1)}\right]
\frac{\partial^2}{\partial z_1^2}\right.\nonumber\\[6pt]
&&+\,i(k_x+k_x')\frac{\sigma\xi'(x_1)}{w(x_1)}\left[\frac{1}{2}+
z_1\frac{\partial}{\partial z_1}\right]\nonumber\\[6pt]
&&-\left.\frac{\sigma^2{\xi'}^2(x_1)}{w^2(x_1)}\left[\frac{1}{4}+
2z_1\, \frac{\partial}{\partial z_1}+
z_1^2\frac{\partial^2}{\partial z_1^2}\right]\right\}
\nonumber\\[6pt]
&&\times \exp\left(ik'_xx_1\right)\frac{\sin(k'_zz_1)}{k'_z}.
\label{1RB-Xi-exactF}
\end{eqnarray}
This equation has a peculiar structure very useful in further
analysis. The kernel written in this form also consists of three
groups of terms, however, now they represent different scattering
mechanisms.

Since we are interested in the averaged Green's function, we have
to calculate the binary correlator of $\Xi(k_x,k'_x)$. Therefore,
to avoid very cumbersome calculations, it is reasonable to make a
simplification of $\Xi(k_x,k'_x)$ that does not destroy the
Hermitian structure of each group of terms. To this end, and
taking into account that our interest is in the typical case of
small surface corrugations ($\sigma\ll d$), we can do the
following: expand the factor $1-d^2/w^2(x_1)\approx
2\sigma\xi(x_1)/d$ in the first term of Eq.~(\ref{1RB-Xi-exactF})
and put $w(x_1)\approx d$ in all the others. In such a way, we get
a suitable {\it approximate} expression for the surface scattering
potential,
\begin{eqnarray}
&&\Xi(k_x,k'_x)\approx\int_{-\infty}^{\infty}dx_1\int_0^ddz_1\,
\exp\left(-ik_xx_1\right)\frac{\sin(k_zz_1)}{k_z}\nonumber\\[6pt]
&&\times\left\{\frac{2\sigma\xi(x_1)}{d}
\frac{\partial^2}{\partial z_1^2}+
i(k_x+k_x')\frac{\sigma\xi'(x_1)}{d}\left[\frac{1}{2}+
z_1\frac{\partial}{\partial z_1}\right]\right.\nonumber\\[6pt]
&&-\left.\frac{\sigma^2{\xi'}^2(x_1)}{d^2}\left[\frac{1}{4}+2z_1\,
\frac{\partial}{\partial z_1}+ z_1^2\frac{\partial^2}{\partial
z_1^2}\right]\right\}\nonumber\\[6pt]
&&\times \exp\left(ik'_xx_1\right)\frac{\sin(k'_zz_1)}{k'_z}.
\label{1RB-Xi-Approx}
\end{eqnarray}

The expression (\ref{1RB-Xi-Approx}) contains one term that
depends on the {\it amplitude} of the roughness profile
$\sigma\xi(x_1)$, and two groups of terms that depend on the
roughness {\it gradient} $\sigma\xi'(x_1)$ and the roughness {\it
square-gradient} $\sigma^2{\xi'}^2(x_1)$ terms, respectively. An
interesting point to mention is that the last group, due to its
proportionality to $\sigma^2$, was neglected in all previous
studies of transport properties of surface-disordered waveguides.
However, as we show below, the scattering due to these terms has
the properties very different from those described by other terms,
and should be properly taken into account.

In order to proceed further, we assert that the kernel can be
written as the sum of its average, $\langle\Xi(k_x,k'_x)\rangle$,
and fluctuating, $\widetilde\Xi(k_x,k'_x)$, parts, i.e.,
\begin{equation}\label{Xi=<Xi>+tildeXi}
\Xi(k_x,k'_x)=\langle\Xi(k_x,k'_x)\rangle+\widetilde\Xi(k_x,k'_x).
\end{equation}
It can be tested that the average, $\langle\Xi(k_x,k'_x)\rangle$,
contributes only to the real part $\gamma_n$ of the complex
renormalization $\delta k_n$ of the lengthwise wave number $k_n$
(see Eq.~(\ref{k+dk})), and therefore, does not change static
transport properties of the surface disordered waveguide. Thus, we
will omit it because our interest is in the attenuation length
$L_n$, not in $\gamma_n$.

From Eq.~(\ref{1RB-Xi-Approx}) one can see that all terms with the
linear dependence on $\sigma$ have zero mean-value. Therefore, the
average part $\langle\Xi(k_x,k'_x)\rangle$ of the kernel
$\Xi(k_x,k'_x)$ is associated only with the group of terms
containing $\sigma^2{\xi'}^2(x_1)$. To extract their fluctuating
contribution, we should subtract from ${\xi'}^2(x_1)$ its mean
value, $\langle{\xi'}^2(x_1)\rangle$. In such a way, we introduce
the following zero-mean-valued operator
\begin{equation}\label{V-def}
\widehat{\cal V}(x)={\xi'}^2(x)-\langle{\xi'}^2(x)\rangle\,,\qquad
\langle\widehat{\cal V}(x)\rangle=0.
\end{equation}
The operator $\widehat{\cal V}(x)$ plays a special role in our
further consideration. In accordance with the Gaussian nature of
the surface-profile function $\xi(x)$, the operator $\widehat{\cal
V}(x)$ is uncorrelated with both $\xi(x)$ and $\xi'(x)$,
\begin{equation}\label{Corr}
\langle\xi(x)\widehat{\cal V}(x')\rangle=0\,,\qquad
\langle\xi'(x)\widehat{\cal V}(x')\rangle=0.
\end{equation}
Its pair correlator is given by
\begin{equation}\label{V-corr}
\langle{\widehat{\cal V}(x)}{\widehat{\cal V}(x')}\rangle=
2\langle\xi'(x)\xi'(x')\rangle^2=2{{\cal W}''}^2(x-x').
\end{equation}
Since we use the $k_x$-representation, it is worthwhile to define
the Fourier transform of the operator ${\widehat{\cal V}(x)}$,
\begin{eqnarray}
V(k_x)&=&\int_{-\infty}^{\infty}dx\exp(-ik_xx)\widehat{\cal V}(x).
\label{FT-V}
\end{eqnarray}
Also, we need the correlator of its Fourier transform,
\begin{equation} \label{FT-VV}
\langle V(k_x)V(k_x')\rangle=4\pi\delta(k_x+k_x')\,T(k_x).
\end{equation}
Here the roughness-square-gradient power (RSGP) spectrum $T(k_x)$
is
\begin{equation} \label{T-def}
T(k_x)=\int_{-\infty}^{\infty}dx\exp{(-ik_xx)}{{\cal W}''}^2(x).
\end{equation}
One should stress that on the one hand, through the integration by
parts, the power spectrum of the roughness gradients $\xi'(x)$ can
be reduced to the RHP spectrum $W(k_x)$. On the other hand, it is
not possible to do the same for the RSGP spectrum $T(k_x)$. This
very fact reflects a highly non-trivial role of the
square-gradient scattering (SGS), giving rise to its competition
with the well known scattering mechanism, in spite of the seeming
smallness of the term $\sigma^2{\xi'}^2(x)$.

With the introduction of $V(k_x)$ we are ready to explicitly write
down the fluctuating part of the total scattering potential,
\begin{equation}
\widetilde \Xi(k_x,k_x')=\Xi_1(k_x,k_x')+\Xi_2(k_x,k_x').
\label{Xi-tilde-def}
\end{equation}
The first summand is associated with the first and second terms in
Eq.~(\ref{1RB-Xi-Approx}),
\begin{subequations}\label{1RB-Xi1}
\begin{eqnarray}
&&\Xi_1(k_x,k_x')=\int_{-\infty}^{\infty}dx_1\int_0^ddz_1
\exp(-ik_xx_1)\,\frac{\sin(k_zz_1)}{k_z} \nonumber\\[6pt]
&&\times\left\{\frac{2\sigma\xi(x_1)}{d}\frac{\partial^2}{\partial
z_1^2}+i(k_x+k_x')\frac{\sigma\xi'(x_1)}{d}\left[\frac{1}{2}+
z_1\frac{\partial}{\partial z_1}\right]\right\}\nonumber\\[6pt]
&&\times\exp(ik_x'x_1)\,\frac{\sin(k_z'z_1)}{k_z'}
\label{1RB-Xi1-def}\\[6pt]
&&=-\sigma[A(k_x,k_x')+B(k_x,k_x')]\widetilde\xi(k_x-k_x').
\label{1RB-Xi1-exp}
\end{eqnarray}
\end{subequations}
The second summand is related to the third term in
Eq.~(\ref{1RB-Xi-Approx}), with ${\xi'}^2(x_1)$ being replaced by
$\widehat{\cal V}(x_1)$,
\begin{subequations}\label{1RB-Xi2}
\begin{eqnarray}
&&\Xi_2(k_x,k_x')=-\int_{-\infty}^{\infty}dx_1\int_0^ddz_1
\exp(-ik_xx_1)\frac{\sin(k_zz_1)}{k_z} \nonumber\\[6pt]
&&\times\frac{\sigma^2\widehat{\cal V}(x_1)}{d^2}
\left[\frac{1}{4}+2z_1\,\frac{\partial}{\partial z_1}+
z_1^2\frac{\partial^2}{\partial z_1^2}\right]\nonumber\\[6pt]
&&\times\exp(ik_x'x_1)\frac{\sin(k_z'z_1)}{k_z'}
\label{1RB-Xi2-def}\\[6pt]
&&=\sigma^2D(k_x,k_x')V(k_x-k_x').
\label{1RB-Xi2-exp}
\end{eqnarray}
\end{subequations}

In Eq.~(\ref{1RB-Xi1}) the quantity $\widetilde\xi(k_x)$ is the
Fourier transform of the function $\xi(x)$,
\begin{eqnarray}
\widetilde\xi(k_x)&=&\int_{-\infty}^{\infty}dx\,\exp(-ik_xx)\,\xi(x).
\label{FT-xi}
\end{eqnarray}
Also, we have introduced the following quantities in the first
summand:
\begin{subequations}\label{1RB-A}
\begin{eqnarray}
&&A(k_x,k_x')=\frac{2k_z'}{k_zd}\int_{0}^{d}dz\,\sin(k_zz)
\sin(k_z'z),\qquad\qquad\label{1RB-A-def}\\[6pt]
&&A(\pm k_n,\pm k_{n'})=\delta_{nn'}\,.\label{1RB-A-exp}
\end{eqnarray}
\end{subequations}
\begin{subequations}\label{1RB-B}
\begin{eqnarray}
&&B(k_x,k_x')=\frac{k_x^2-k_x'^2}{d}\int_0^ddz
\,\frac{\sin(k_zz)}{k_z}\nonumber\\[6pt]
&&\times\left[\frac{1}{2}+z\frac{\partial}{\partial z}\right]
\frac{\sin(k_z'z)}{k_z'}\,,\label{1RB-B-def}\\[6pt]
&&B(\pm k_n,\pm k_{n'})=\cos[\pi(n-n')](1-\delta_{nn'}).
\qquad\qquad\label{1RB-B-exp}
\end{eqnarray}
\end{subequations}
In the last expression we have used the equality
\begin{equation}\label{ecl}
k_x^2-{k_x'}^2=-(k_z^2-{k_z'}^2)
\end{equation}
that directly follows from the energy conservation
$k^2=k_x^2+k_z^2={k_x'}^2+{k_z'}^2$ (see the definition
(\ref{kz})). And finally, in the second summand the following
factors appeared,
\begin{subequations}\label{1RB-D1}
\begin{eqnarray}
&&D(k_x,k_x')=-\frac{1}{d^2}\int_0^ddz
\,\frac{\sin(k_zz)}{k_z}\nonumber\\[6pt]
&&\times\left[\frac{1}{4}+ 2z\,\frac{\partial}{\partial z}+
z^2\frac{\partial^2}{\partial z^2}\right]
\frac{\sin(k_z'z)}{k_z'}\,,\label{1RB-D-def}\\[6pt]
&&D(\pm k_n,\pm k_{n'})=\frac{d}{2}\left[\frac{1}{3}+
\frac{1}{(2\pi n)^2}\right]\delta_{nn'}
\nonumber\\[6pt]
&&+\frac{2 d}{\pi^2}\,\frac{n^2+n'^2}{(n^2-n'^2)^2}\,
\cos[\pi(n-n')](1-\delta_{nn'}).\qquad\qquad\label{1RB-D-exp}
\end{eqnarray}
\end{subequations}

The operator $\widetilde\Xi(k_x,k_x')$ written as the sum
(\ref{Xi-tilde-def}) of specially designed terms, has a very
convenient form. First, both terms are chosen to have zero
average. Second, since the operators $\xi(x)$ and $\widehat{\cal
V}(x)$ do not correlate with each other (see Eq.~(\ref{Corr})),
their Fourier transforms are also uncorrelated,
\begin{equation}\label{FT-Corr}
\langle\tilde\xi(k_x)V(k_x')\rangle=0.
\end{equation}
Due to the condition (\ref{FT-Corr}), the scattering potentials
$\Xi_1(k_x,k_x')$ and $\Xi_2(k_x,k_x')$ are also uncorrelated.
However, they have the following autocorrelators
\begin{subequations}\label{Xi1-corr-Q1}
\begin{eqnarray} \label{Xi1-corr}
&&\langle\Xi_1(k_x,q_x)\Xi_1(q_x,k_x')\rangle=\nonumber\\
&&\qquad\qquad\qquad 2\pi\delta(k_x-k_x')\,Q_1(k_x,q_x),\\[6pt]
\label{1RB-Q1}
&&Q_1(k_x,q_x)=\sigma^2W(k_x-q_x)\,[A(k_x,q_x)+B(k_x,q_x)]\quad
\nonumber\\
&&\qquad\qquad\times[A(q_x,k_x)+B(q_x,k_x)]\,.
\end{eqnarray}
\end{subequations}
\begin{subequations}\label{Xi2-corr-Q2}
\begin{eqnarray}\label{Xi2-corr}
&&\langle\Xi_2(k_x,q_x)\Xi_2(q_x,k_x')\rangle=\nonumber\\
&&\qquad\qquad\qquad 2\pi\delta(k_x-k_x')\,Q_2(k_x,q_x),\qquad\\[6pt]
\label{1RB-Q2}
&&Q_2(k_x,q_x)=2\sigma^4T(k_x-q_x)\nonumber\\
&&\qquad\qquad\times D(k_x,q_x)D(q_x,k_x)\,.
\end{eqnarray}
\end{subequations}
As a result, the correlator of the fluctuating scattering
potential (\ref{Xi-tilde-def}) in the $k_x$-representation, is as
follows
\begin{subequations}\label{Xi-tilde-corr}
\begin{eqnarray}
&&\langle\widetilde\Xi(k_x,q_x)\widetilde\Xi(q_x,k_x')\rangle=
2\pi\delta(k_x-k_x')Q(k_x,q_x),\qquad\quad\label{Xi-Q}\\[6pt]
&&Q(k_x,q_x)=Q_1(k_x,q_x)+Q_2(k_x,q_x)\,.\label{Q}
\end{eqnarray}
\end{subequations}

It is now possible to perform an appropriate perturbative
averaging of the Dyson equation (\ref{g-DE}), and at the same
time, to separate a relative contribution of the SGS mechanism
from the total scattering process.

\section{Average Green's Function}
\label{sec-1RB-AverageGF}

Now we are in a position to replace the problem for the random
Green's function ${\cal G}(x,x';z,z')$ with the problem for the
Green's function $\langle{\cal G}(x,x';z,z')\rangle$ averaged over
the surface disorder. It is evident that $\langle{\cal
G}(x,x';z,z')\rangle$ is governed by Eq.~(\ref{G-pole}) with the
average pole factor $\langle g(k_x,k_x')\rangle$ instead of the
random one. To perform the averaging of Eq.~(\ref{g-DE}) with
$\Xi(k_x,k_x')$ given by Eq.~(\ref{Xi-tilde-def}) and obtain
$\langle g(k_x,k_x')\rangle$, we can apply one of the standard and
well known perturbative methods. For example, it can be the
diagrammatic approach developed for surface disordered
systems~\cite{BFb79}, as well as the technique developed in
Ref.~\onlinecite{McGM84}. Both of the methods allow one to develop
the consistent perturbative approach with respect to the
scattering potential, that takes adequately into account the {\it
multiple scattering} from the corrugated boundary. Then, we come
to the following result
\begin{eqnarray}
&&\langle g(k_x,k_x')\rangle=g_0(k_x,k_x')
\nonumber\\[6pt]
&+&\int_{-\infty}^{\infty}\frac{dq_xdq_x'dq_x''dq_x'''}
{(2\pi)^4}\,g_0(k_x,q_x)
\nonumber\\[6pt]
&\times&\langle\widetilde\Xi(q_x,q_x')g_0(q_x',q_x'')\widetilde
\Xi(q_x'',q_x''')\rangle\langle g(q_x''',k_x')\rangle.\qquad\
\label{<g>-Dyson}
\end{eqnarray}
Taking into account the presence of delta-functions in the
definitions (\ref{g0-kk'}) for the unperturbed pole factor and in
Eq.~(\ref{Xi-Q}) for the scattering potential, we can take
explicitly the integrals over $q_x$, $q_x''$ and $q_x'''$ in the
second term. After that, Eq.~(\ref{<g>-Dyson}) becomes an
algebraic one. Its solution has the form,
\begin{subequations}\label{<g>}
\begin{eqnarray}
\langle g(k_x,k_x')\rangle&=&2\pi\delta(k_x-k_x')\,g(k_x),
\label{<g>-kk'}\\[6pt]
g(k_x)&=&\left[g_0^{-1}(k_x)-M(k_x)\right]^{-1}.
\label{g-def}
\end{eqnarray}
\end{subequations}
The quantity $M(k_x)$ is called the {\it self-energy or mass operator}.
It is described by the formula
\begin{equation}\label{M-def}
M(k_x)=\int_{-\infty}^{\infty}\frac{dq_x}{2\pi}\,Q(k_x,q_x)\,g_0(q_x).
\end{equation}

One should remind that when obtaining Eq.~(\ref{<g>-Dyson}) we
have omitted the average part of the kernel $\Xi(k_x,k_x')$. The
motivation to discard it, as was stated after
Eq.~(\ref{Xi=<Xi>+tildeXi}), arises because its contribution to
the renormalization of the lengthwise wave number $k_n$ is real,
i.e. it contributes only to the quantity $\gamma_n$ (see
Eq.~(\ref{k+dk})). Besides this, another contribution to
$\gamma_n$ arises due to the real part of the self-energy. For
this reason, we can expand the average pole factor $g(k_x)$ in
series of partial fractions and retain only the imaginary part of
the self-energy $M(k_n)$,
\begin{eqnarray}
g(k_x)&\to&\sum_{n=1}^{N_d}\frac{(\pi n/d)^2}{k_nd}
\left(\frac{1}{k_x+k_n+i/2L_n}\right. \nonumber\\[6pt]
&&-\left.\frac{1}{k_x-k_n-i/2L_n}\right).
\label{g-pf}
\end{eqnarray}
This expression completely corresponds to the representation
(\ref{g0-pf}) for the unperturbed pole factor. It is
suitable for further evaluation of the average Green's function.

The quantity $L_n$ is the {\it total} wave {\it attenuation
length} or electron {\it mean-free-path} of the $n$-th conducting
mode. It describes the scattering from $n$-th mode into all
possible propagating modes. This quantity is determined by the
imaginary part of the self-energy $M(k_n)$,
\begin{subequations}\label{Ln-gen}
\begin{eqnarray}
L_n^{-1}&=&-2\frac{(\pi n/d)^2}{k_nd}\,\Im M(k_n)
\label{Ln-def}\\[6pt]
&=&\frac{(\pi n/d)^2}{k_nd}\,\sum_{n'=1}^{N_d}
\frac{(\pi n'/d)^2}{k_{n'}d}
\nonumber\\[6pt]
&&\times\left[Q(k_n,-k_{n'})+Q(k_n,+k_{n'})\right].
\label{Ln-Q}
\end{eqnarray}
\end{subequations}

Let us substitute Eqs.~(\ref{<g>-kk'}) and (\ref{g-pf}) into
Eq.~(\ref{G-pole}) and perform straightforward calculations of the
integrals with the use of the delta-function and over the poles of
$g(k_x)$. As a result, we find the average Green's function,
\begin{equation}\label{<G>lineG}
\langle{\cal G}(x,x';z,z')\rangle=\overline{\cal G}(|x-x'|;z,z'),
\end{equation}
in the efficient representation via canonical Fourier series in
the normal waveguide modes,
\begin{eqnarray}
&&\overline{\cal G}(|x-x'|;z,z')=\sum_{n=1}^{N_d}
\sin\left(\frac{\pi nz}{d}\right)
\sin\left(\frac{\pi nz'}{d}\right)
\nonumber\\[6pt]
&&\times\frac{\exp(ik_n|x-x'|)}{ik_nd}\,
\exp\left(-\frac{|x-x'|}{2L_n}\right).\label{avGF}
\end{eqnarray}

\section{Attenuation Length Analysis}
\label{sec-1RB-Ln-An}

In view of Eq.~(\ref{Q}), the general expression (\ref{Ln-gen})
for the inverse attenuation length shows that in the problem under
consideration this quantity consists of two terms,
\begin{equation}\label{1RB-Ln-sum}
\frac{1}{L_n}=\frac{1}{L^{(1)}_n}+\frac{1}{L^{(2)}_n}.
\end{equation}
These terms descend from different mechanisms of the surface
scattering. The first attenuation length $L^{(1)}_n$ is related
with the RHP spectrum, $W(k_x)$, through the expression for
$Q_1(k_x,k'_x)$. In accordance with Eqs.~(\ref{1RB-Q1}),
(\ref{1RB-A-exp}) and (\ref{1RB-B-exp}), it is given by
\begin{eqnarray}
\frac{1}{L^{(1)}_n}&=&
\sigma^2\frac{(\pi n/d)^2}{k_nd}\,\sum_{n'=1}^{N_d}
\frac{(\pi n'/d)^2}{k_{n'}d}\nonumber\\[6pt]
&&\times\left[ W(k_n+k_{n'})+W(k_n-k_{n'})\right].
\qquad\qquad\label{1RB-Ln1-def}
\end{eqnarray}
Its {\it diagonal term} is formed by the {\it amplitude
scattering} (AS) and the {\it off-diagonal terms} result from the
{\it gradient scattering} (GS). These two mechanisms of surface
scattering are due to the corresponding terms in the expression
for $\Xi_1(k_x,k_x')$ (see Eq.~(\ref{1RB-Xi1-def})), i.e., the
former from the term depending on the {\it amplitude} of the
roughness profile $\sigma\xi(x)$, and the latter from the terms
depending on the roughness {\it gradient} $\sigma\xi'(x)$. The
expression (\ref{1RB-Ln1-def}) exactly coincides with that
previously obtained by various methods (see, e.g.,
Ref.~\onlinecite{BFb79}).

The second attenuation length $L^{(2)}_n$  related to RSGP
spectrum through $Q_2(k_x,k_x')$, is associated solely with the
SGS mechanism due to the operator $\widehat{\cal V}(x)$ (see
Eq.~(\ref{1RB-Xi2-def})). In accordance with Eqs.~(\ref{1RB-Q2})
and (\ref{1RB-D-exp}), it is described by
\begin{equation}\label{1RB-LnG-mode}
\frac{1}{L^{(2)}_n}=\sum_{n'=1}^{N_d}\frac{1}{L^{(2)}_{nn'}}.
\end{equation}
Here its diagonal term controlling  the wave scattering inside the
mode ({\it intramode scattering}) is written as
\begin{eqnarray}
\frac{1}{L^{(2)}_{nn}}&=&\frac{\sigma^4}{2}\frac{(\pi
n/d)^4}{k_n^2}\,\left[\frac{1}{3}+\frac{1}{(2\pi n)^2}\right]^2
\nonumber\\[6pt]
&\times&\left[T(2k_n)+T(0)\right].\label{1RB-LnnG}
\end{eqnarray}
The off-diagonal partial scattering length $L^{(2)}_{n\neq n'}$
that describes the {\it intermode scattering} (from $n$-th mode
to $n'$-th one, $n\neq n'$), is
\begin{eqnarray}
\frac{1}{L^{(2)}_{n\neq n'}}&=&\frac{8\sigma^4}{\pi^4}\frac{(\pi
n/d)^2}{k_n}\,\frac{(\pi n'/d)^2}{k_{n'}}\,
\frac{(n^2+n'^2)^2}{(n^2-n'^2)^4}
\nonumber\\[6pt]
&\times&\left[T(k_n+k_{n'})+T(k_n-k_{n'})\right].
\label{1RB-Lnn'2}
\end{eqnarray}
To the best of our knowledge, in the surface-scattering problem
for multimode waveguides the operator ${\hat{\cal V}(x)}$ was
never taken into account, and, as a result, the second attenuation
length, or SGS length, $L^{(2)}_n$, was missed in previous
studies.

Now we list the simplifications that have been made in deriving
Eqs.~(\ref{1RB-Ln-sum}) -- (\ref{1RB-Lnn'2}) for the attenuation
length $L_n$. First, the proper self-energy (\ref{M-def}) in the
Dyson-type equation for the average Green's function has been
obtained within the second-order approximation in the perturbation
potential. In terms of the diagrammatic technique this is similar
to the ``simple vortex" or, the same, Bourret
approximation~\cite{Bourret62} that contains the binary correlator
$Q(k_x,q_x)$ of the surface-scattering potential and the
unperturbed pole factor $g_0(q_x)$. To find out the conditions of
applicability for this approach, we have used the ideas proposed
in the book~\onlinecite{RytKrTat89}. More specifically, we
substitute into the self-energy (\ref{M-def}) the average pole
factor (\ref{g-def}) instead of the unperturbed one. This trick is
equivalent to the summation of an infinite subsequence of diagrams
in the exact expansion of the self-energy in powers of the
scattering potential. The analysis shows that in the Dyson-type
equation the new (and more general) self-energy can be reduced to
ours, if the channel broadening $1/2L_n$ is much less than the
unperturbed quantum wave number $k_n$ and the variation scale
$R_c^{-1}$ of the RHP and RSGP spectra, i.e., when $2k_nL_n\gg1$
and $R_c\ll2L_n$.

Second, in order to extract the inverse attenuation length from
the self-energy (\ref{M-def}), we have changed the lengthwise wave
number $k_x$ by its unperturbed value $k_n$. This change is
justified if the surface-induced broadening $1/2L_n$ can be
neglected in comparison with $R_c^{-1}$ and the spacing
$|k_n-k_{n\pm1}|\approx|\partial k_n/\partial n|$ between
neighboring quantum wave numbers. Now we take into account that
$|\partial k_n/\partial n|\sim\Lambda_n^{-1}$, where $\Lambda_n$
is the distance between two successive reflections of the $n$-th
mode from the rough boundary. Therefore, the use of $k_n$ instead
of $k_x$ in the argument of the self-energy is valid under the
conditions $R_c\ll2L_n$ and $\Lambda_n\ll2L_n$.

Thus, we come to three requirements: $R_c\ll2L_n$,
$\Lambda_n\ll2L_n$, and $2k_nL_n\gg1$. Due to the obvious
relationship $k_n\Lambda_n\gtrsim1$, the last inequality is a
direct consequence of the second one, and one can conclude that
the domain of applicability for our results is restricted by two
independent criteria of {\it weak surface scattering},
\begin{equation}\label{WS-Ln}
\Lambda_n=2k_nd/(\pi n/d) \ll2L_n, \qquad R_c\ll2L_n.
\end{equation}
They imply that the wave is weakly attenuated on both the
correlation length $R_c$ and the cycle length $\Lambda_n$. From
the analysis performed above, it becomes clear that expressions
(\ref{1RB-Ln1-def}) and (\ref{1RB-LnG-mode}) -- (\ref{1RB-Lnn'2})
represent main contributions from the substantially distinct
surface-scattering mechanisms: AS+GS and SGS. In particular, the
corrections that are proportional to $\sigma^4$, originated from
the next order of approximation in the {\it amplitude} and {\it
gradient} terms of the surface-scattering potential, can not
compete with the main contribution (\ref{1RB-Ln1-def}) under the
conditions (\ref{WS-Ln}). On the contrary, the {\it
square-gradient} terms give rise to the $\sigma^4$-terms in
Eqs.~(\ref{1RB-LnnG}) -- (\ref{1RB-Lnn'2}), which should not be
neglected due to a specific dependence on the correlation length
$R_c$. Note that Eq.~(\ref{WS-Ln}) implicitly includes the
requirement $\sigma\ll d$ for the surface corrugations be small in
height, that has been employed in the section~\ref{sec-1RB-Dyson}
when deriving the explicit form (\ref{1RB-Xi-Approx}) for the
surface scattering potential $\Xi(k_x,k_x')$.

We are now in a position to analyze the attenuation length $L_n$.
For convenience, in our further analysis, we deal with the
attenuation lengths in the form of the dimensionless quantities
$\Lambda_n/2L_n$, $\Lambda_n/2L_n^{(1)}$ and
$\Lambda_n/2L_n^{(2)}$. Since $L^{(1)}_n$ and $L^{(2)}_n$ depend
on as many as four dimensionless parameters $(k\sigma)^2$, $kR_c$,
$kd/\pi$, and $n$, the complete analysis appears to be quite
complicated. For this reason, below we restrict ourselves by the
analysis of the interplay between $L^{(1)}_n$ and $L^{(2)}_n$ as a
function of the dimensionless correlation length $kR_c$, for
different values of $(k\sigma)^2$ and mode index $n$. In the
analysis, we consider a multimode waveguide, i.e., the situation
when the number of propagating modes is large, $N_d\approx
kd/\pi\gg 1$. From the physical point of view, two types of rough
surfaces seem to be the most important. Surfaces of the first type
contain a small-scale roughness of the ``white noise'' kind when
$kR_c\ll 1$. For the second type, the waveguide surface consists
of large-scale random corrugations when $kR_c \gg 1$. We shall
develop our analysis for this two types of surfaces.

\subsection{Small-Scale Roughness}
\label{subsec-1RB-Ln-SSR}

Let us start with a relatively simple and widely used case of {\it
a small-scale boundary perturbation}, when $kR_c\ll 1$ and the
surface roughness can be regarded as a delta-correlated random
process with the correlator ${\cal W}(x-x')\approx
W(0)\delta(x-x')$ and constant power spectrum $W(k_x)\approx
W(0)\sim R_c$. Taking into account the evident relationship
$k\Lambda_n\gtrsim 1$, one can get the following inequalities to
specify this case
\begin{equation}\label{SSR-def}
kR_c\ll 1\lesssim k\Lambda_n.
\end{equation}
It is necessary to underline that in the regime of small-scale
roughness (\ref{SSR-def}) the second of the weak-scattering
conditions in Eq.(\ref{WS-Ln}) is not so restrictive as the first
one, and directly stems from it, $R_c\ll\Lambda_n\ll2L_n$.

In Eqs.~(\ref{1RB-Ln1-def}), (\ref{1RB-LnnG}) and
(\ref{1RB-Lnn'2}) for the attenuation lengths the argument of the
correlators $W(k_x)$ and $T(k_x)$ turns out to be much less than
the scale of their decrease $R_c^{-1}$ under the conditions
(\ref{SSR-def}). Therefore, for any term in the summation over
$n'$ the argument can be taken as zero.

Therefore, the first attenuation length is determined as follows,
\begin{subequations}\label{1RB-Ln1-SSR}
\begin{eqnarray}
&&\frac{\Lambda_n}{2L^{(1)}_n}\approx
2(k\sigma)^2\frac{n}{kd/\pi}\frac{W(0)}{k}\sum_{n'=1}^{N_d}
\frac{(\pi n'/d)^2}{k_{n'}d}\qquad
\label{1RB-Ln1-SSR-Sum}\\[6pt]
&&\approx (k\sigma)^2\frac{n}{kd/\pi}\frac{kW(0)}{2}
\label{1RB-Ln1-SSR-W}
\end{eqnarray}
\end{subequations}
Due to a large number of the conducting modes $N_d\approx
kd/\pi\gg 1$, we can change the summation over $n'$ by
integration. In this way one can obtain Eq.~(\ref{1RB-Ln1-SSR-W})
from Eq.~(\ref{1RB-Ln1-SSR-Sum}). In order to correctly estimate
the result, one can take into account the formula
\begin{equation}\label{W(0)}
W(0)=\int_{-\infty}^{\infty}dx{\cal W}(x)
=2R_c\int_{0}^{\infty}d\rho{\cal W}(R_c\rho),
\end{equation}
which directly follows from the definition (\ref{FT-W}) for the
Fourier transform $W(k_x)$ of the binary correlator ${\cal W}(x)$.
The function ${\cal W}(R_c\rho)$ is the dimensionless correlator
of the dimensionless variable $\rho$, with the scale of decrease
of the order of one. As a result, the function ${\cal W}(R_c\rho)$
does not depend on $R_c$. Therefore, the integral over $\rho$
entering Eq.~(\ref{W(0)}) is a positive constant of the order of
unity. For example, $W(0)=\sqrt{2\pi}R_c$ and the integral is
$\sqrt{\pi/2}$ in the case of Gaussian correlations (see
Eq.~(\ref{WFT-Gaus})).

For the SGS length we have
\begin{subequations}\label{1RB-Ln2-SSR}
\begin{eqnarray}
&&\frac{\Lambda_n}{2L^{(2)}_n}\approx
\pi^2\frac{(k\sigma)^4}{k_nd} \frac{n^3}{(kd/\pi)}\frac{T(0)}{k^3}
\nonumber\\[6pt]
&&\times\Bigg\{\left[\frac{1}{3}+\frac{1}{(2\pi n)^2}\right]^2
+\frac{16k_nd}{\pi^4(\pi n/d)^2}
\nonumber\\[6pt]
&&\times\left(\sum_{n'=1}^{n-1}+\sum_{n'=n+1}^{N_d}\right)
\frac{(\pi n'/d)^2}{k_{n'}d}\frac{(n^2+n'^2)^2}{(n^2-n'^2)^4}\Bigg\}
\qquad\label{1RB-Ln2-SSR-Sum}\\[6pt]
&&\approx\frac{\pi^2}{4}\frac{(k\sigma)^4}{k_nd}
\frac{n^3}{(kd/\pi)}\frac{T(0)}{k^3}. \label{1RB-Ln2-SSR-NT}
\end{eqnarray}
\end{subequations}
In Eq.~(\ref{1RB-Ln2-SSR-Sum}) every term in the sum rapidly
decreases with an increase of absolute value of $\Delta n=n-n'$.
This can be seen by making use the following estimate,
\begin{equation}\label{factorDeltan}
\frac{(n^2+{n'}^2)^2}{(n^2-{n'}^2)^4}\approx\frac{1}{4(\Delta
n)^4}\qquad\mbox{for}\quad n\gg|\Delta n|.
\end{equation}
The fast decrease of the factor is supported by direct
calculations, see Table~\ref{cuadro_factorDeltan}. Therefore, the
sum in Eq.~(\ref{1RB-Ln2-SSR-Sum}) can be well evaluated by two
terms with $n'=n\pm1$. For simplicity, in
Eq.~(\ref{1RB-Ln2-SSR-NT}) we assume $N_d\gg n\gg 1$, and replace
the curly braces by factor $1/4$.

\begin{table}
\caption{\label{cuadro_factorDeltan} Evaluation of
Eq.~(\ref{factorDeltan})}
\begin{center}
\begin{tabular}{|r|*{8}{|r}|}\hline
$n\diagdown\Delta n$&-4&-3&-2&-1&1&2&3&4\\
\hline \hline
1& 0{.}002& 0{.}006& 0{.}024& 0{.}31&&&& \\
2&$\cdots$& 0{.}004& 0{.}019& 0{.}27& 0{.}31& && \\
3&$\cdots$& $\cdots$& 0{.}018& 0{.}26& 0{.}27& 0{.}024& & \\
4&$\cdots$& $\cdots$& $\cdots$& 0{.}26& 0{.}26& 0{.}019& 0{.}006& \\
5&$\cdots$& $\cdots$& $\cdots$& $\cdots$& 0{.}26& 0{.}018&0{.}004& 0{.}002 \\
$\vdots$&&&&&&&& \\
\hline
\end{tabular}
\end{center}
\end{table}

The explicit form for $T(0)$ directly follows from the definition
(\ref{T-def}) for the correlator $T(k_x)$,
\begin{eqnarray}\label{T(0)}
T(0)&=&\int_{-\infty}^{\infty}dx{{\cal W}''}^2(x)
\nonumber\\[6pt]
&=&R_c^{-3}\int_{-\infty}^\infty d\rho \left[\frac{d^2{\cal
W}(R_c\rho)}{d\rho^2}\right]^2.
\end{eqnarray}
If the roughness correlations are of the Gaussian form, then
according to Eq.~(\ref{T-Gaus-expl}), we have
$T(0)=3\sqrt{\pi}/4R_c^3$ and the integral over $\rho$ entering
Eq.~(\ref{T(0)}) is equal to $3\sqrt{\pi}/4$.

The relationship between the attenuation lengths is expressed by
\begin{equation}\label{L1/L2-SSR}
\frac{L^{(1)}_n}{L^{(2)}_n}\sim\,(k\sigma)^2
\frac{n^2}{k_nd}\,(kR_c)^{-4}.
\end{equation}
According to substantially different behavior of the quantities
$\Lambda_n/2L^{(1)}_n$ and $\Lambda_n/2L^{(2)}_n$ with respect to
$kR_c$, it becomes clear that they must intersect at the crossing
point $(kR_c)_{cr}$. If the crossing point falls onto the present
region of small-scale roughness ($kR_c\ll1$), its dependence on
the model parameters is obtained by equating to one the expression
(\ref{L1/L2-SSR}),
\begin{equation}\label{cp-SSR}
(kR_c)_{cr}^2\sim (k\sigma)n/\sqrt{k_nd}.
\end{equation}
To the left from this point $(kR_c)_{cr}$ the SGS length prevails,
$L^{(2)}_n\ll L^{(1)}_n$. To its right the main contribution is
due to the first attenuation length, $L^{(1)}_n\ll L^{(2)}_n$. The
expression (\ref{cp-SSR}) shows that the crossing point is smaller
for smaller values of the dimensionless roughness height
$k\sigma$, as well as for smaller mode indices $n$, or for larger
values of the mode parameter $kd/\pi$.

\subsection{Large-Scale Roughness: Weak Correlations}
\label{subsec-1RB-Ln-WC}

The intermediate situation arises when the correlation length
$R_c$ becomes much larger than the wave length $2\pi/k$, but still
remains much less than the cycle length $\Lambda_n$,
\begin{equation}\label{LSR-WC-def}
1\ll kR_c\ll k\Lambda_n.
\end{equation}
As before, the first of the weak-scattering conditions
(\ref{WS-Ln}) is the most restrictive, i.e., we get
$R_c\ll\Lambda_n\ll2L_n$.

Since the distance $\Lambda_n$ is larger than the correlation
length $R_c$, successive reflections of the waves from the rough
surface are weakly correlated. Meantime, the distance between
neighboring wave numbers $k_n$ and $k_{n\pm1}$ is much smaller
than the variation scale $R_c^{-1}$ of the correlators $W(k_x)$
and $T(k_x)$,
\begin{equation}\label{kn-WC}
|k_n-k_{n\pm1}|\approx |\partial k_n/\partial n|=
2\pi\Lambda_n^{-1}\ll R_c^{-1}.
\end{equation}
This implies that the correlators $W(k_x)$ and $T(k_x)$ are smooth
functions of the summation index $n'$. Therefore, the sum in the
expression (\ref{1RB-Ln1-def}) for the first attenuation length
$L^{(1)}_n$ can be substituted by the integral,
\begin{subequations}\label{1RB-Ln1-WC}
\begin{eqnarray}
\label{1RB-Ln1-WC-W}
&&\frac{\Lambda_n}{2L^{(1)}_n}\approx \sigma^2\left(\frac{\pi
n}{d}\right)\,\int_{0}^{N_d} dn'\frac{(\pi n'/d)^2}{k_{n'}d}
\nonumber\\[6pt]
&&\times\left[W(k_n+k_{n'})+W(k_n-k_{n'})\right] \\[6pt]
\label{1RB-Ln1-WC-fin} &&=\frac{\sigma^2}{\pi}\left(\frac{\pi
n}{d}\right)\,\int_{-k}^{k}
dk_x\sqrt{k^2-k_x^2}\,W(k_n-k_x).\qquad\qquad
\end{eqnarray}
\end{subequations}
Eq.~(\ref{1RB-Ln1-WC}) shows that the first attenuation length is
contributed by scattering of a given propagating $n$-th mode into
all other propagating modes. Note that to obtain this asymptotic
result we have used only the condition of week correlations,
$R_c\ll\Lambda_n$. Therefore, Eq.~(\ref{1RB-Ln1-WC}) provides the
reduction to Eq.~(\ref{1RB-Ln1-SSR}) for small-scale corrugations,
$kR_c\ll 1$. In the case of large-scale roughness, $kR_c\gg 1$,
the formula (\ref{1RB-Ln1-WC}) obtained for $L^{(1)}_n$ allows
further simplifications as was done in Ref.~\onlinecite{BFb79}.

In contrast to $L^{(1)}_n$, due to a rapidly decaying factor
(\ref{factorDeltan}), the SGS length $L^{(2)}_n$ can be still
described by keeping tree terms only, $n'=n,n\pm 1$, in the sum in
Eq.~(\ref{1RB-LnG-mode}). Taking into account the estimate
(\ref{kn-WC}), for the case $N_d\gg n\gg 1$ one can write down,
\begin{subequations}\label{1RB-Ln2-WC}
\begin{eqnarray}
&&\frac{\Lambda_n}{2L^{(2)}_n}\approx\frac{\pi^2}{2}
\frac{(k\sigma)^4}{k_nd}\frac{n^3}{(kd/\pi)}\frac{T(0)+T(2k_n)}{k^3}
\nonumber\\[6pt]
&&\times\Bigg\{\left[\frac{1}{3}+\frac{1}{(2\pi n)^2}\right]^2
+\frac{8}{\pi^4}\Bigg\}
\label{1RB-Ln2-WC-NT}\\[6pt]
&&\approx\frac{\pi^2}{8}\frac{(k\sigma)^4}{k_nd}
\frac{n^3}{(kd/\pi)}\frac{T(0)+T(2k_n)}{k^3}.
\qquad\quad\label{1RB-Ln2-WC-fin}
\end{eqnarray}
\end{subequations}
In the final expression (\ref{1RB-Ln2-WC-fin}), we have replaced
the curly braces from Eq.~(\ref{1RB-Ln2-WC-NT}) by the factor
$1/4$. Naturally, at small-scale corrugations, $kR_c\ll 1$, the
obtained result (\ref{1RB-Ln2-WC}) passes into
Eq.~(\ref{1RB-Ln2-SSR}). For the large-scale roughness, when
$kR_c\gg 1$, one should use Eq.~(\ref{1RB-Ln2-WC}) because of
arbitrary value of the parameter $k_nR_c$. We do not consider here
this case in detail due to its intermediate character.

\subsection{Large-Scale Roughness: Strong Correlations}
\label{subsec-1RB-Ln-SC}

In the other extreme case, the correlation length $R_c$ is very
large not only in comparison with the wave length $2\pi/k$, but
also in comparison with the cycle length $\Lambda_n$,
\begin{equation}\label{LSR-SC-def}
1\lesssim k\Lambda_n\ll kR_c.
\end{equation}
In this case the number of wave reflections over the correlation
length $R_c$ is large. Therefore, the successive reflections are
strongly correlated to each other.

Under the inequalities (\ref{LSR-SC-def}) the second of the
weak-scattering conditions (\ref{WS-Ln}) is the most restrictive,
therefore, the condition of applicability reads as
\begin{equation}\label{domain-SC}
\Lambda_n\ll R_c\ll2L_n.
\end{equation}
The latter requirement determines the upper limit for the value of
the correlation length $R_c$.

Due to Eq.~(\ref{LSR-SC-def}), the distance between neighboring
wave numbers $k_n$ and $k_{n\pm1}$ turns out to be much larger
than the variation scale $R_c^{-1}$ of the correlators $W(k_x)$
and $T(k_x)$,
\begin{equation}\label{kn-SC}
|k_n-k_{n\pm1}|\approx|\partial k_n/\partial
n|=2\pi\Lambda_n^{-1}\gg R_c^{-1}.
\end{equation}
This indicates that the probability of the intermode ($n'\neq n$)
transitions is exponentially small and the attenuation lengths are
mainly formed by the incoherent intramode ($n'=n$) scattering.
Formally, at strong correlations (\ref{LSR-SC-def}) the
correlators $W(k_x)$ and $T(k_x)$ are sharpest functions of the
summation index $n'$. In the sums of Eqs.~(\ref{1RB-Ln1-def}) and
(\ref{1RB-LnG-mode}) for the attenuation lengths the main
contribution is due to the diagonal terms with $n'=n$, in which
$W(2k_n)$ and $T(2k_n)$ can be neglected in comparison with $W(0)$
and $T(0)$.

Thus, the first attenuation length reads
\begin{equation}\label{1RB-Ln1-SC}
\frac{\Lambda_n}{2L^{(1)}_n}\approx
\frac{(k\sigma)^2}{k_nd}\frac{n^3}{(kd/\pi)^3}\,kW(0).
\end{equation}
Correspondingly, for the SGS length one gets,
\begin{equation}\label{1RB-Ln2-SC}
\frac{\Lambda_n}{2L^{(2)}_n}\approx \frac{\pi^2}{2}
\frac{(k\sigma)^4}{k_nd}\frac{n^3}{(kd/\pi)}\frac{T(0)}{k^3}
\left[\frac{1}{3}+\frac{1}{(2\pi n)^2}\right]^2
\end{equation}

The ratio of the first attenuation length to the second one can be
presented as
\begin{equation}\label{L1/L2-SC<<1}
\frac{L^{(1)}_n}{L^{(2)}_n}\sim\,\frac{\Lambda_n}{2L_n^{(1)}}\,
\left(\frac{\Lambda_n}{R_c}\right)^5(k_n \Lambda_n)^{-2}\ll1.
\end{equation}
According to the first inequality in Eq.~(\ref{WS-Ln}), to the
condition (\ref{LSR-SC-def}) of the strong-correlations and to the
evident relationship $k_n\Lambda_n\gtrsim1$, we can see that the
amplitude scattering length always prevails over the SGS length
within the interval of strong correlations. For this reason, in
the condition of applicability (\ref{domain-SC}) one should
substitute $L_n$ by $L^{(1)}_n$. This allows one to arrive at the
inequalities in the explicit form,
\begin{equation}\label{VR-SC}
k\Lambda_n \ll kR_c \ll (k\Lambda_n) (k\sigma)^{-1} (kd/\pi n).
\end{equation}

\subsection{Numerical Analysis}
\label{subsec-1RB-Ln-Num}

In this subsection we perform the numerical analysis of the
scattering lengths for the case when the random surface profile
$\xi(x)$ has the Gaussian binary correlator,
\begin{equation}\label{corr-Gauss}
{\cal W}(x)=\exp(-x^2/2R_c^2).
\end{equation}
Then the RHP spectrum (\ref{FT-W}) is given by
\begin{equation}\label{WFT-Gaus}
W(k_x)=\sqrt{2\pi}\,R_c\,\exp(-k_x^2R_c^2/2).
\end{equation}
The RSGP spectrum $T(k_x)$ defined by Eq.~(\ref{T-def})), can be
explicitly presented as
\begin{eqnarray}
T(k_x)
&=&\frac{\sqrt{\pi}}{16R_c^3}\left[(k_xR_c)^4-4(k_xR_c)^2+12\right]
\nonumber\\[6pt]
&\times&\exp\left[-(k_xR_c)^2/4\right]\,.
\label{T-Gaus-expl}
\end{eqnarray}

\begin{figure}[t]
\includegraphics[angle=270,width=\columnwidth]{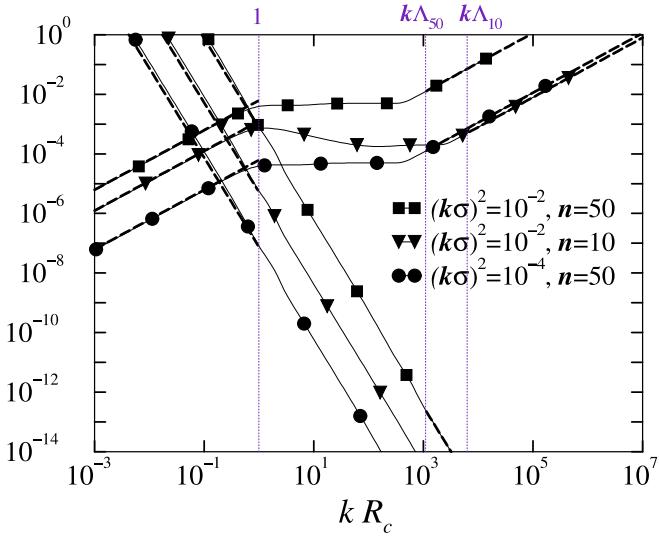}
\caption{\label{fig:1RB_approx_al_L1L2_kRc} Plots of
$\Lambda_n/2L_n^{(1)}$ (increasing curves) and
$\Lambda_n/2L_n^{(2)}$ (decreasing curves) vs. $kR_c$ for
$kd/\pi=100.5$. Dashed lines show the corresponding asymptotic
expressions.}
\end{figure}

In Fig.~\ref{fig:1RB_approx_al_L1L2_kRc} we display separately the
behavior of $\Lambda_n/2L_n^{(1)}$ and $\Lambda_n/2L_n^{(2)}$, as
a function of the dimensionless correlation parameter $kR_c$,
comparing them with the analytically obtained asymptotics. The
dashed lines are used to plot the asymptotics (\ref{1RB-Ln1-SSR})
and (\ref{1RB-Ln2-SSR}) for the region (\ref{SSR-def}) of
small-scale roughness ($kR_c\ll1\lesssim k\Lambda_n$), and
expressions (\ref{1RB-Ln1-SC}) and (\ref{1RB-Ln2-SC}) for the
region (\ref{LSR-SC-def}) of large-scale roughness with strong
correlations ($1\lesssim k\Lambda_n\ll kR_c$). As one can see, for
these two regions numerical data for both lengths are quite well
described by the corresponding asymptotic expressions.

The curves $\Lambda_n/2L_n^{(1)}$ clearly manifest two transition
points: between the regions of small-scale and large-scale
corrugations at $kR_c\sim1$, and between weak and strong
correlations at $kR_c\sim k\Lambda_{n}$. As follows from
Eq.~(\ref{1RB-Ln1-def}), the inverse value of the first
attenuation length typically increases with an increase of $kR_c$.
Specifically, within the interval of the small-scale roughness
($kR_c\ll 1\lesssim k\Lambda_n$) we have $1/L^{(1)}_n\propto
kR_c$. Then, within the intermediate region of large-scale
roughness with weak correlations where $1\ll kR_c\ll k\Lambda_n$,
the increase of $\Lambda_n/2L_n^{(1)}$ slows down (see the curves
with the parameters $(k\sigma)^2=10^{-2},10^{-4}$; $n=50$), or can
even be replaced by the decrease for some values of the model
parameters (see the curve with the parameter
$(k\sigma)^2=10^{-2}$, $n=10$). Within these two regions that are
unified under the condition $kR_c\ll k\Lambda_n$, the quantity
$1/L^{(1)}_n$ is determined by both AS and GS mechanisms (AS+GS).
Finally, for large-scale roughness and strong correlations
($1\lesssim k\Lambda_n\ll kR_c$) the value of $1/L^{(1)}_n$ again
begins to increase linearly with $kR_c$. Here $1/L^{(1)}_n$ is
associated solely with the AS mechanism because the main
contribution to the asymptotic~(\ref{1RB-Ln1-SC}) is due to the
diagonal term in the sum (\ref{1RB-Ln1-def}) .

In contrast with $1/L^{(1)}_n$, the inverse SGS length $1/L^{(2)}_n$
reveals a monotonous decrease as
the parameter $kR_c$ increases. At small ($kR_c\ll 1\lesssim
k\Lambda_n$) and extremely large ($1\lesssim k\Lambda_n\ll kR_c$)
values of $kR_c$, this decrease obeys the law
$1/L^{(2)}_n\propto(kR_c)^{-3}$, due to $T(0)\sim R_c^{-3}$.

In Fig.~\ref{fig:1RB_approx_al_L1L2_kRc} we can also see the
crossover from the SGS to AS+GS that is characterized by the
crossing point between $\Lambda_n/2L_n^{(2)}$ and
$\Lambda_n/2L_n^{(1)}$. The crossing point for two curves with the
parameters $(k\sigma)^2=10^{-4}$ and $n=50$ is very close to that
for two curves with the parameters $(k\sigma)^2=10^{-2}$ and
$n=10$. Approximately, both crossing points are
\begin{equation}\label{xpoint-num}
(kR_c)_{cr}\sim 0.2.
\end{equation}
They are located well inside the interval of small-scale
roughness, and their values are in an agreement with the
asymptotic expression (\ref{cp-SSR}). Two curves corresponding to
the parameters $(k\sigma)^2=10^{-2}$ and $n=50$, have crossing
point in the transition region $kR_c\sim1$ between small- and
large-scale corrugations.

\begin{figure}[t]
\includegraphics[angle=270,width=\columnwidth]{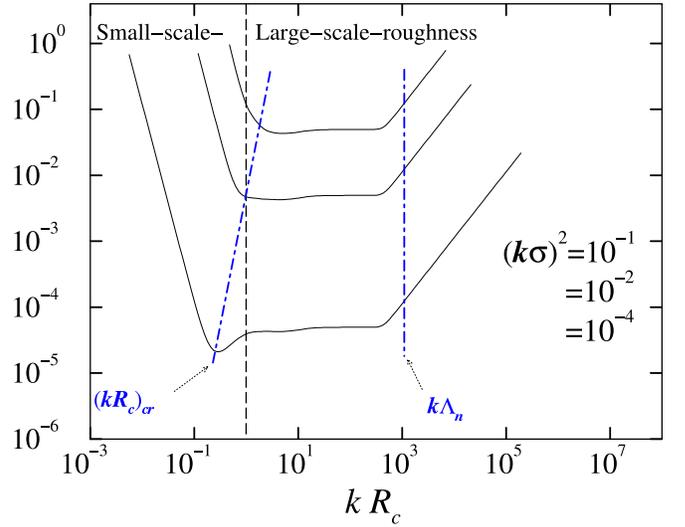}
\caption{\label{fig:1RB_al_L1L2_kRc} Plots of $\Lambda_n/2L_n$
versus $kR_c$ for $k d/\pi=100.5$, $n=50$, and different values of
$(k\sigma)^2$. The dot-dashed lines labelled `$(kR_c)_{cr}$' and
`$k\Lambda_n$', indicate the transition between the regions
dominated by the SGS, AS+GS and AS mechanisms (from the left to
the right). The dashed line at $kR_c=1$ marks the transition
between the regions of small-scale- and large-scale-roughness.}
\end{figure}

\begin{figure}[t]
\includegraphics[angle=270,width=\columnwidth]{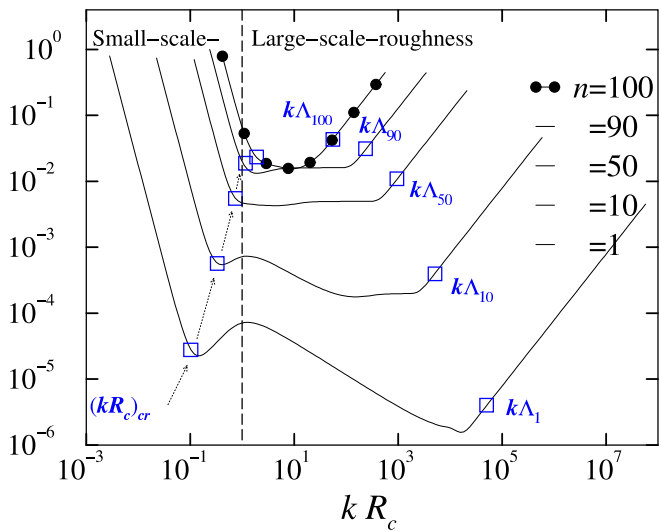}
\caption{\label{fig:1RB_approx_al_L1L2_n_kRc} Plots of
$\Lambda_n/2L_n$ versus $kR_c$ for $k d/\pi=100.5$,
$(k\sigma)^2=10^{-2}$, and different values of the mode index $n$.
The sets of square symbols labelled `$(kR_c)_{cr}$' and
`$k\Lambda_n$' play the same role as dot-dashed lines in
Fig.~\ref{fig:1RB_al_L1L2_kRc}. }
\end{figure}

In the following two Figures we display the dependence of
$\Lambda_n/2L_n$ as a function of $kR_c$. The curves are plotted
starting from the values of $kR_c$ for which
$\Lambda_n/2L_n^{(2)}=1$, according to the first condition in
(\ref{WS-Ln}). Taking into account the second condition
restricting the maximal value of $kR_c$, we plot every curve
within the range where $R_c<2L^{(1)}_n$. Based upon the
description of the Fig.~\ref{fig:1RB_approx_al_L1L2_kRc}, the
identification of each scattering mechanism dominating in the
corresponding regions becomes simple. One can see that the curves
in Figs.~\ref{fig:1RB_al_L1L2_kRc} and
\ref{fig:1RB_approx_al_L1L2_n_kRc} experience firstly the
crossover from the SGS to the AS+GS, and after, from the AS+GS to
AS. We outline both transitions with the labels `$(kR_c)_{cr}$'
and `$k\Lambda_n$', respectively.

In Fig.~\ref{fig:1RB_al_L1L2_kRc} we show the dependence of
$\Lambda_n/2L_n$ on $kR_c$ for three values of the parameter
$(k\sigma)^2$ (two of them $(k\sigma)^2=10^{-4},10^{-2}$
correspond to those used in
Fig.~\ref{fig:1RB_approx_al_L1L2_kRc}). The curve with
$(k\sigma)^2=10^{-4}$ has the crossing point $(kR_c)_{cr}$ of the
value (\ref{xpoint-num}) located within the interval of
small-scale roughness. The crossover reveals a small dip centered
at $(kR_c)_{cr}$. The curve obeys the asymptotic behavior
$(kR_c)^{-3}$ to the left from $(kR_c)_{cr}$ due to the main
contribution from $\Lambda_n/2L_n^{(2)}$. After, the quantity
$\Lambda_n/2L_n^{(1)}$ becomes dominating in the sum
(\ref{1RB-Ln-sum}), therefore, the curve begins to rise. Firstly,
the linear dependence on $kR_c$ on the right deep-side (where
$kR_c<1$) is replaced with a smoother one (for $kR_c>1$). Finally,
for $R_c>\Lambda_n$ (strong correlations) the linear dependence
restores.

The crossing points of the second and third curves with
$(k\sigma)^2=10^{-2},10^{-1}$ have values of the order of unit,
$(kR_c)_{cr}\sim 1$. Here the total attenuation length $L_n$
within the whole small-scale region is formed solely by the SGS
length $L_n^{(2)}$. In full agreement with Eq.~(\ref{cp-SSR}) the
presented curves display that the smaller the parameter
$(k\sigma)^2$ the smaller the value of the crossing point
$(kR_c)_{cr}$.

To visualize the dependence on the mode index $n$, in
Fig.~\ref{fig:1RB_approx_al_L1L2_n_kRc} we plot $\Lambda_n/2L_n$
with the parameter $n$ spanning the range of the propagating
modes. One can see that for the mode index $n=1$ the crossing
point $(kR_c)_{cr}\sim0.1$, the curve with $n=10$ has
$(kR_c)_{cr}$ given by Eq.~(\ref{xpoint-num}). For the rest of the
values of $n$ the crossover occur at $(kR_c)_{cr}\sim1.0$. We note
that the smaller the channel index $n$, the smaller the value of
the crossing point $(kR_c)_{cr}$. This fact is in full agreement
with Eq.~(\ref{cp-SSR}). The squares labelled with `$k\Lambda_n$',
show the transition points between the AS+GS and AS mechanisms (or
between the weak- and strong-correlations of the surface
roughness). They are $k \Lambda_{1} \approx 63500$, $k
\Lambda_{10} \approx 6315$, $k \Lambda_{50} \approx 1100$, $k
\Lambda_{90} \approx 310$ and $k \Lambda_{100} \approx 60$.

Finally, note that for all curves in our figures the roughness
{\it height} is small, $\sigma/d\ll1$. Furthermore, for the
amplitude- and gradient-dominated scattering (to the right from
the point $(kR_c)_{cr}$ where $\Lambda_n/2L^{(1)}_{n}$ mainly
contributes), the average corrugation {\it slope} is also small
for all data, $\sigma/R_c\ll1$. The roughness slope remains to be
small at the crossing points too, but increases to their left with
the decrease of $kR_c$. As a result, to the left from the crossing
point where the square-gradient term $\Lambda_n/2L^{(2)}_{n}$
prevails, the slope reaches the values of the order one, or even
larger.

\section{Conclusion}
\label{Sec-Conclusion}

In this paper we investigated the wave/electron scattering in
multi-mode surface-corrugated waveguides of quasi-1D geometry. For
this kind of waveguides, we have discovered a new {\it
square-gradient scattering} (SGS) mechanism that is different from
the previously studied ones. This mechanism arises due to specific
square-gradient terms in the Hamiltonian describing the surface
scattering, that are related to the {\it roughness
square-gradient} power (RSGP) spectrum $T(k_x)$. To compare with,
the well known scattering mechanisms, the {\it amplitude} (AS) and
the {\it gradient} (GS) ones, are both determined by  the {\it
roughness-height} power (RHP) spectrum $W(k_x)$, only. Since the
SGS mechanism is independent from the others, one can define two
attenuation lengths, the known length $L_n^{(1)}$ and the SGS
length $L_n^{(2)}$. Both contribute to the total attenuation
length (or, the same, electron mean free path) $L_n$ according to
Eq.~(\ref{1RB-Ln-sum}).

The roughness-height $W(k_x)$ and square-gradient $T(k_x)$ power
spectra have very different dependencies on the roughness
correlation length $R_c$. This provides the substantially
different behavior of the corresponding scattering lengths in
dependence of the model parameters. Specifically, the inverse
value of the first attenuation length $1/L^{(1)}_n$ typically
increases, while the inverse value of the SGS length $1/L^{(2)}_n$
decreases, with an increase of the parameter $kR_c$. Therefore,
the curves displaying these quantities intersect upon the increase
of the dimensionless correlation length $kR_c$, and the crossover
from the SGS to AS+GS occurs. To the left from the crossing point
$(kR_c)_{cr}$ the SGS length prevails over the first attenuation
length, $L^{(2)}_n\ll L^{(1)}_n$. To the right from $(kR_c)_{cr}$,
the first attenuation length mainly contributes to the scattering
process, $L^{(1)}_n\ll L^{(2)}_n$. If the crossing point
$(kR_c)_{cr}$ falls into the interval (\ref{SSR-def}) of
small-scale surface corrugations, it obeys the law (\ref{cp-SSR}).

As we have shown, at any fixed value of the root-mean-square
roughness height $\sigma$, one can indicate the region of small
values of the correlation length $R_c$ where the new attenuation
length $L^{(2)}_{n}$ predominates over the known length
$L^{(1)}_{n}$. This predominance arises in spite of the fact that
$1/L^{(1)}_{n}$ is proportional to $\sigma^2$ while
$1/L^{(2)}_{n}$ is proportional to $\sigma^4$.

In the large-scale roughness regime where the first attenuation
length mainly contributes, $L^{(1)}_n\ll L^{(2)}_n$, one can
observe two different behaviors of $1/L_n^{(1)}$. In the interval
of weak correlations (\ref{LSR-WC-def}) the dependence of
$1/L_n^{(1)}$ on $kR_c$ is quite complicated, due to the
coexistence of both the AS and GS mechanisms. However in the
region of strong correlations (\ref{LSR-SC-def}), because the AS
stands alone, we have much simpler behavior, $1/L^{(1)}_n\propto
kR_c$. It is remarkable that the SGS mechanism prevails in the
widely discussed region of a small-scale boundary perturbation,
$kR_c\ll1$, where the surface roughness is typically described via
the white-noise potential.

\begin{acknowledgments}
This research was supported by Consejo Nacional de Ciencia y
Tecnolog\'{\i}a (CONACYT, M\'exico) under the grant No. 43730,
and by the Universidad Aut\'onoma de Puebla (BUAP, M\'exico)
under the grant 5/G/ING/05.
\end{acknowledgments}



\end{document}